\documentclass[reprint,aps,pdftex,dvipsnames]{revtex4-2}

\usepackage{graphicx}
\usepackage{dcolumn}
\usepackage{bm}
\usepackage{tikz}          
\usepackage{placeins}      
\usepackage{graphicx}          
\usepackage{natbib}

\usepackage{wrapfig}
\usepackage{hyperref}

\usepackage{float}

\usepackage[version=4]{mhchem}
\usepackage{subfiles}
\usepackage{braket}
\usepackage{amsmath}
\usepackage{amssymb}
\usepackage{physics}
\usepackage{mathtools}
\usepackage[super]{nth}
\usepackage{siunitx}
\usepackage{tikz}

\usepackage{tikz}
\usepackage{enumitem}

\usepackage{lipsum}

\usepackage{multirow}
\usepackage{tabularx}
\usepackage{tabularray}
\usepackage{booktabs}
\usepackage{rotating}
\usepackage{cases}
\usepackage{tensor}
\usepackage{xargs}                    
\usepackage{xurl}
\usepackage{xfrac}

\begin{document}

\preprint{APS/123-QED}
\title{Low-depth measurement-based deterministic quantum state preparation}

\author{Roselyn Nmaju, Fiona Speirits, Sarah Croke}
\date{\today}

\begin{abstract}
We present a low-depth amplitude encoding method for arbitrary quantum state preparation. Building on the foundation of an existing divide-and-conquer algorithm, we propose a method to disentangle the ancilla qubits from the final state. Our method is measurement-based but deterministic, and offers an alternative approach to existing state preparation algorithms. It has circuit depth $\Theta(n^2)$ and requires $\Theta(2^n)$ ancilla qubits. The resulting spacetime allocation is $\Theta(2^n)$, which is optimal. Furthermore, we go beyond scaling arguments to precisely quantify the resources needed for any $n$. We illustrate our method through detailed worked examples of both a ``dense'' state and a W-state. Finally, we demonstrate how to combine this method with existing methods to construct hybrid algorithms with varying space and circuit depth complexities.
\end{abstract}

\maketitle

\section{Introduction}
Quantum state preparation is a fundamental task in quantum computation: in order to process quantum or classical data, it is first necessary to load the data as a quantum state. This is especially (but not exclusively) pertinent for machine learning applications~\cite{HHL_2009,lloyd2014quantum,kerenidis2016quantum}, where the complexity of data loading can negate any quantum advantage over classical or quantum-inspired classical techniques~\cite{tang2019quantum,chia2022sampling}, if not carefully accounted for~\cite{aaronson2015read,tang2021quantum}. 

Without the help of ancilla qubits, the circuit depth needed to prepare an arbitrary quantum state scales exponentially with the number of qubits, $n$~\cite{mottonen2005transformation,plesch2011quantum,shende2005synthesis,sun2023asymptotically}. Recent work has explored the tradeoff between time (circuit depth) and space (ancilla qubits)~\cite{Araujo_2021,zhang2021low,zhang2022,clader2022quantum,sun2023asymptotically,Yuan2023optimalcontrolled,Gui_2024}. It is known that an $n$ qubit state may be prepared with a circuit depth polynomial in $n$ with the use of a number of ancilla qubits that scales superpolynomially with $n$. Optimal strategies achieve $\Theta(n)$ depth with $\Theta(2^n/n)$ ancilla qubits~\cite{Yuan2023optimalcontrolled,Gui_2024}. For sparse states, with $d$ non-zero entries, this can be further reduced to circuit depth $\order{\log (n d) }$ and $\order{n d \log d}$ ancilla qubits~\cite{zhang2022,10.1109/DAC18074.2021.9586240,Malvetti_2021}.

In this paper, we consider and extend the divide-and-conquer state preparation method of Araujo \emph{et al.}~\cite{Araujo_2021}. This method, outlined below, aims to prepare an $n$-qubit state by successively combining smaller states pairwise to build up the desired state. However, it requires exponentially many ancilla qubits, which will remain entangled with the data qubits. In this work, we show how to disentangle the data qubits from the ancilla qubits at each step. The resulting method requires only single-qubit measurements with classical feed-forward, and is deterministic, despite being measurement based. Combined with the original divide-and-conquer proposal, it may be applied recursively to produce a low-depth state preparation algorithm, needing $\Theta(n^2)$ depth and $\Theta(2^n)$ ancilla qubits to produce an $n$-qubit state. The recently introduced spacetime allocation metric~\cite{Gui_2024}, a measure of total resources needed for a protocol, scales as $\Theta(2^n)$ for our protocol, which is optimal.

Previous work~\cite{zhang2021low} also considered concatenation of states and measurement-based disentangling, but the resulting protocol was probabilistic. The probability of success could be made $\order{1}$ through repetition, with the resulting protocol requiring either a run time or number of qubits scaling as $\order{N^2} = \order{2^{2n}}$. An alternative approach uses a flag mechanism and controlled unitaries to disentangle ancilla qubits~\cite{clader2022quantum,Gui_2024}. The method proposed here thus has different features to existing methods, and gives an alternative approach to the problem of arbitrary quantum state preparation.

We demonstrate our method through detailed examples, and precisely quantify the resource requirements. Finally, we introduce the combine-and-conquer method to create a hybrid protocol combining the divide-and-conquer method with existing encoding methods~\cite{mottonen2005transformation}. In this approach, smaller sub-circuits may be constructed from any alternative encoding method and used as the building blocks for the divide-and-conquer method. 

\section{Preliminaries}

We begin with some background and notation needed in the rest of the paper.
The state preparation problem may be presented as follows: given the list $\vec{x}= [x_0, x_1, x_2, ..., x_{N-1}]$, where $x_i$ are complex numbers, and $\sum_i |x_i|^2 = 1$, the goal is to produce the state:
\begin{align} 
    \ket{\Psi} &= x_0 \ket{00...0} + x_1 \ket{00...1} + ... + x_{N-1} \ket{11...1} \nonumber \\
    &= \sum_{i=0}^{N-1} x_i \ket{i}, \label{eq: amplitude_encoding}
\end{align}
where $\ket{i}$ is a computational basis state representing the integer $i$ in binary notation. This is referred to as \textit{amplitude encoding} of the classical vector $\vec{x}$, and an $N$-dimensional vector may be encoded in $n=\log_2(N)$ qubits in this way.

We assume that the list $\vec{x}$ is of length $N=2^n$; if this is not the case, then $0's$ may be appended until it is. We also assume that $\vec{x}$ is normalised; if this is not the case, we can divide by an overall normalisation factor as a preliminary step before applying the procedures described.

We will also assume familiarity with single-qubit gates and make use of $\hat{R}_y(\theta)$ and $\hat{R}_z(\phi)$ rotations about the $y$-axis and $z$-axis in the Bloch sphere by an angle $\theta$, $\phi$ respectively:
\begin{align}
    \hat{R}_y(\theta) &= \exp{-i \theta \hat{\sigma}_y/2} = \cos \frac{\theta}{2} \hat{I} -i \sin \frac{\theta}{2} \hat{\sigma}_y \label{eq:y rotation},\\
    \hat{R}_z(\phi) &= \exp{-i \phi \hat{\sigma}_z/2} = \cos \frac{\phi}{2} \hat{I} -i \sin \frac{\phi}{2} \hat{\sigma}_z.
\end{align}
An arbitrary single-qubit state, represented by the vector $\vec{x} = [x_0,x_1]$, can be parameterised as:
\begin{align}
    x_0 &= \cos (\frac{\theta}{2}), \\
    x_1 &= e^{i \phi} \sin (\frac{\theta}{2}). \label{eq:theta1 value}
\end{align}
Starting in state $\ket{0}$, a $y$-rotation is therefore sufficient to set the amplitude, followed by a $z$-rotation to set the phase:
\begin{align}
    \ket{\Psi}_{\rm 1q} &= x_0 \ket{0} + x_1 \ket{1}, \nonumber \\
    &= \cos (\theta/2) \ket{0} + e^{i \phi} \sin (\theta/2 )\ket{1}, \nonumber \\
    &= e^{i \phi/2} \hat{R}_z(\phi) \hat{R}_y (\theta) \ket{0}.
\end{align}

In the remainder of the paper, we will restrict ourselves to real, positive vectors $\vec{x}$ for clarity, although the extension to complex vectors is straightforward.

It is useful to visualise the state encoding algorithms described below as traversing a binary tree. We introduce this structure and the notation used to label it here. An $ N=2^n$-dimensional state may be represented by a tree with levels $l=\{0,\ldots,n\}$, and with $N$ leaf nodes. Nodes are associated with computational basis states; moving to the left (right) along the tree corresponds to appending a $\ket{0}$ ($\ket{1}$) respectively to the current computational basis state. Edges between nodes are labelled with weights such that: i) each path through the tree to a leaf node leads to a unique $n$-qubit computational basis state and, ii) the product of the weights of all the edges traversed in a given path gives the amplitude associated with that computational basis state. The state represented by such a tree is therefore a superposition over computational basis states, with amplitudes given by the weights of the corresponding paths.
\begin{figure}
    \centering
    \includegraphics[width=0.9\linewidth]{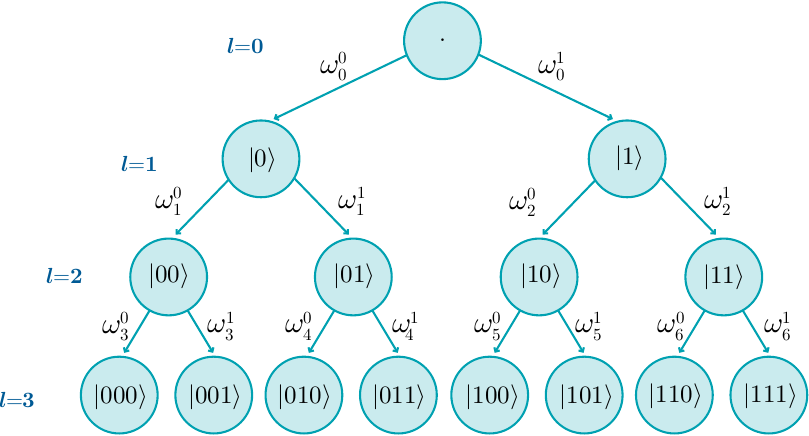}
    \caption{Recursive binary tree structure with weights $\omega_f^j$ picked up at each level for each state.}
    \label{fig:weighted binary tree}
\end{figure}
Each node of the tree is labelled with a subscript $f(l,p) = 2^{l} - 1 + p$, where $l$ is the level of the binary tree and $p$ is the position across the tree within each level, counting from $0$ to $2^{l} -1$. We define $l=0$ to be the root node, and the leaf nodes correspond to $l=n$. The weights associated with the edges originating at a given node $f$ are labelled $\omega_f^0$, $\omega_f^1$ respectively, where the superscript is the basis state that will be picked up by traversing that direction down the binary tree.

For example, starting at the root node $(0)$, traversing the path weighted by $\omega_0^1$ takes the null state at $(0)$ and appends to it the state $\ket{1}$, which is at node $(2)$. This is illustrated for $N=8$ in Fig. \ref{fig:weighted binary tree}. With this notation, the node at level $l$ and position $p$ is associated with the $l$-qubit computational basis state $\ket{p_1 p_2 \ldots p_{l}}$ where the string $\{p_1 p_2 \ldots p_{l}\}$ is the binary representation of $p$. The weight $\omega_f^0$ is then the total amplitude for the $(l+1)$th qubit to be in state $\ket{0}$, conditional on the first $l$ qubits being in states $\ket{p_1 p_2 \ldots p_{l}}$.

To calculate the weights $\omega_f^0$, $\omega_f^1$ for a given state with coefficients $\{ x_i \}$, we work backwards up the tree from the leaves to the root: at level $l=n-1$, we can define
\begin{align}\label{eq:omega}
\omega_{f(n-1,p)}^j = \begin{cases} 
          \frac{x_{2p+j}}{\sqrt{\abs{x_{2p}} ^2 + \abs{x_{2p+1}}^2}}, & \sqrt{\abs{x_{2p}} ^2 + \abs{x_{2p+1}}^2} >0, \\\\
          0,  & \sqrt{\abs{x_{2p}} ^2 + \abs{x_{2p+1}}^2}= 0.
       \end{cases}
\end{align}
At lower levels we recursively define vectors $\{ x_i^{(l)} \}$
\begin{equation}\label{eq: amplitude iteration}
x_i^{(l)} = \sqrt{\abs{x_{2i}^{(l+1)}} ^2 + \abs{x_{2i+1}^{(l+1)}} ^2}, \quad l = 0, 1, \ldots n-2 \, ,
\end{equation}
where $x_i^{(n-1)} = x_i$. The weights $\omega_f^0$, $\omega_f^1$ at each level are then defined analogously to Eq.~\eqref{eq:omega}, and it may be verified that the product of the weights associated with any given path gives the correct coefficient $x_i$.

\begin{figure}[h!]
    \centering
    \includegraphics[width=\linewidth]{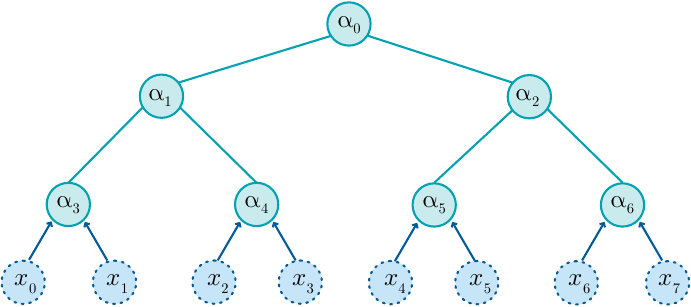}
    \caption{Recursive binary tree structure
    [solid-teal] of a general input vector $\vec{x}$ of size $N = 8$ [dashed-blue].}
    \label{fig:Recursion Tree}
\end{figure}

Finally, we assign an angle $\alpha_f$ to each node. This is defined as the angle of rotation of the gate needed to prepare the single-qubit state:
\begin{equation}\label{eq:nodestate}
    \ket{\phi_f} = \omega_f^0 \ket{0} + \omega_f^1 \ket{1}.
\end{equation}
Following Eq.~\eqref{eq:theta1 value} we therefore define
\begin{equation}\label{eq:alphaangles}
\alpha_f = 2\sin^{-1}\left(\tensor*{\omega}{*_f ^1}\right).
\end{equation}
The resulting simplified binary tree structure, in which nodes are labelled by the angles $\alpha_f$, is shown in Fig.~\ref{fig:Recursion Tree}.

\begin{figure*}[t]
    \centering
    \includegraphics[width=0.9\linewidth]{ 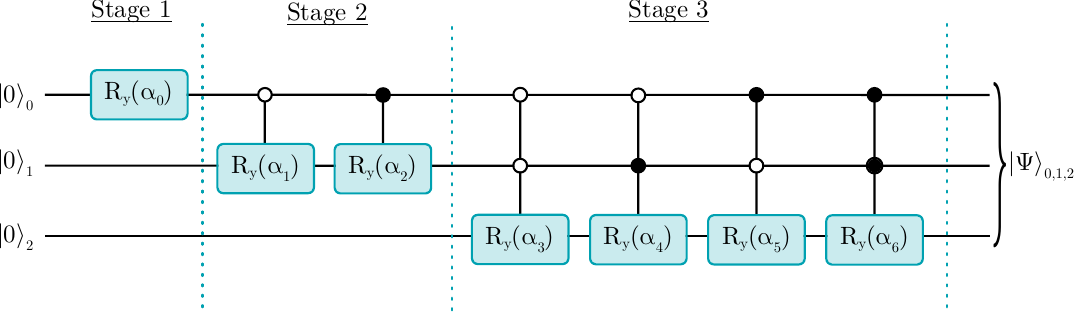}
    \caption{Quantum circuit realising the general state preparation using the time encoding method for an $N=8$ real vector input.}
    \label{fig:N=8 Time Encoding}
\end{figure*}

In the absence of ancilla qubits, the most straight-forward approach to state preparation starts at the root of the tree, preparing a single-qubit state at the first stage. Subsequent stages, moving down through the tree, correspond to conditional rotations on qubit $l+1$, conditional on the states of all previous $l$ qubits. The circuit realising this general state preparation is shown in Fig.~\ref{fig:N=8 Time Encoding}. In general, encoding an $n$-qubit input requires $\order{2^n}$ rotations applied consecutively across $n$ qubits~\cite{mottonen2005transformation,plesch2011quantum}. This exponential scaling in circuit depth, coupled with the increase in the size of the multi-controlled gates, which themselves need to be decomposed into simpler operations, means that the scaling of a time-encoded circuit is unfavourable for use in encoding larger data sets~\cite{PhysRevA.106.042602,9475957}. Recent work therefore has explored the possibility of reducing the circuit depth by using ancilla qubits, trading time for space.

\section{The Divide-and-Conquer Method}\label{sec:  Divide-and-Conquer}
The divide-and-conquer method proposed in~\cite{Araujo_2021} can be understood as traversing a binary tree, following what the authors refer to as a \textit{bottom-up} approach, working up from the leaves to the root~\cite{Araujo_2021, Araujo_2023, zhang2021low}. At the core of the divide-and-conquer protocol is a proposed \textit{state combining technique}, which for any given $k$ takes two $(k-1)$-qubit states along with a single-qubit ancilla and combines these to produce a $k$-qubit state. Note that if we can achieve this for any $k$, then an arbitrary quantum state can be built from the bottom up as the two $(k-1)$-qubit states are themselves built by combining two $(k-2)$-qubit states and so on until it is a matter of creating single-qubit states. Altogether, starting from $2^{n-1}$ single-qubit states at the leaves of the tree, and combining these pairwise at each level as we move up through the binary tree structure, we can create any general state of the form Eq.~\eqref{eq: amplitude_encoding}. This is described in detail in~\cite{Araujo_2021}. Clearly, this requires a number of ancilla qubits which scales exponentially with $n$.

Specifically, suppose that we know how to produce two arbitrary $(k-1)$-qubit states, which we label $\ket{\varphi_0}$ and $\ket{\varphi_1}$, and we wish to prepare the $k$-qubit state
\begin{equation} \label{eq: combiner}
    a \ket{0} \ket{\varphi_0} + b \ket{1} \ket{\varphi_1},    
\end{equation}
where $a, b \in \mathbb{C}$.
The proposition of~\cite{Araujo_2021} is to use the control SWAP (CSWAP) gates for this state-combining step. Starting with the states $\ket{\varphi_0}$, $\ket{\varphi_1}$ prepared in different registers, and a control qubit prepared in state $a \ket{0} + b \ket{1}$, a CSWAP operation (in which each of the $k$ qubits of $\ket{\varphi_0}$ are swapped with the corresponding qubit of $\ket{\varphi_1}$, controlled on the same control qubit) produces the state
\begin{equation}\label{eq: divide_and_conquer}
\begin{split} 
    U_{\mathrm{CSWAP}}((&a\ket{0} + b\ket{1}),\ket{\varphi_0},\ket{\varphi_1}) = \\
   &a \ket{0} \ket{\varphi_0} \ket{\varphi_1} + b \ket{1} \ket{\varphi_1} \ket{\varphi_0}. 
\end{split}
\end{equation}
This is almost the desired state, but not quite, as it remains entangled with an ancilla register (comparing Eq.~\eqref{eq: combiner} with Eq.~\eqref{eq: divide_and_conquer}).

The authors show that by using this module recursively, it is possible to prepare a state of the form
\begin{equation}
    \ket{\Psi}_{\rm dc} = \sum_{i=0}^{N-1} x_i \ket{i} \ket{\psi_i},
\label{eq:ancilla qubits}
\end{equation}
where $\{ \ket{\psi_i} \}$ are orthogonal states of the ancilla qubits. Thus, the probabilities associated with each computational basis state $\ket{i}$ are successfully reproduced, but any phase information in the desired state is lost due to entanglement with the ancilla. Nonetheless, some applications of information encoded in this form are demonstrated in~\cite{Araujo_2021}. In the next section, we propose a method to disentangle the ancilla qubits from the data qubits at each step, thus resulting in a fully general state preparation algorithm.

\section{Disentangling Measurement}

\subsection{Ancilla Register Measurement}

To disentangle the ancilla qubits from the data qubits, we propose performing a measurement on the ancilla register, i.e.~the third register in Eq.~\ref{eq: divide_and_conquer}. If the measurement on the ancilla qubits projects onto pure states, then the ancilla will be decoupled after the measurement. The key idea is to choose a measurement that, in addition, does not alter the relative amplitude of the $\ket{0} \ket{\varphi_0}$, $\ket{1} \ket{\varphi_1}$ components in the data registers (the first two registers) in Eq.~\eqref{eq: divide_and_conquer}. Note that $\ket{\varphi_0}$ and $\ket{\varphi_1}$ in the ancilla register span a two-dimensional subspace, so we need only consider the effect of measurement on this subspace. We therefore wish to construct a basis $\{\ket{\pm}\}$ in this subspace with the property that $|\braket{\pm}{\varphi_0}| = |\braket{\pm}{\varphi_1}|$.

To demonstrate this explicitly, we introduce the relative phases $\delta_\pm$, defined such that
\begin{equation}
    \braket{\pm}{\varphi_0} = e^{i \delta_{\pm}} \braket{\pm}{\varphi_1}.
\label{eq:deltapm}
\end{equation}
Consider now the effect of performing a measurement with this property on the third register of the state Eq.~(\ref{eq: divide_and_conquer}). Applying the projector $\{\ketbra{\pm}{\pm}\}$ to update the state gives:
\begin{equation}
\begin{gathered}
        \left( a \ket{0}\ket{\varphi_0} \right) \ket{\pm} \braket{\pm}{\varphi_1} + b \left( \ket{1}\ket{\varphi_1} \right) \ket{\pm} \braket{\pm}{\varphi_0} \\
        = \braket{\pm}{\varphi_1} (a \ket{0}\ket{\varphi_0} + b e^{i \delta_{\pm}} \ket{1}\ket{\varphi_1}) \ket{\pm},
\end{gathered}
\end{equation}
where we have used Eq.~(\ref{eq:deltapm}). Discarding the ancilla register, which is now in a product state with the other two, and re-normalising, we see that the resulting state for each possible outcome is given by
\begin{equation}
    a \ket{0} \ket{\varphi_0} + b e^{i \delta_\pm}\ket{1} \ket{\varphi_1}.
\end{equation}
The relative phase term $e^{i \delta_{\pm}}$ can be easily corrected with a single-qubit gate: $R_z(\delta_\pm)$ on the control qubit, conditioned classically on the result of measurement. Thus, irrespective of the measurement result, we can recover the desired state. 

It remains to be shown that we can construct a measurement basis with the desired properties \emph{and} perform the measurement efficiently. Assuming, without loss of generality, that the overlap $\braket{\varphi_0}{\varphi_1}$ is real and positive (we can always choose the global phase of either $\ket{\varphi_0}$ or $\ket{\varphi_1}$ to ensure this is true), such a basis is constructed as
\begin{equation}\label{eq:plusminus}
    \ket{\pm} = \frac{\ket{\varphi_0} \pm \ket{\varphi_1}}{\sqrt{2(1 \pm \braket{\varphi_0}{\varphi_1}})}.
\end{equation}
It is readily verified that these states are orthonormal and have the desired properties: $\braket{\pm}{\varphi_0} = \pm \braket{\pm}{\varphi_1}$. 

We conclude with a discussion of the complexity of such a measurement. If $\ket{\varphi_0}$ and $\ket{\varphi_1}$ are single-qubit states, then $\ket{\pm}$ defines a single-qubit measurement, which is readily implemented. However, in the general $(k-1)$-qubit case, the $\ket{\pm}$ states generically will be multipartite states over all $(k-1)$-qubits, and there seems to be no guarantee that the desired measurement can be implemented efficiently. Fortunately, it is known that any two orthogonal states may be perfectly distinguished with only local measurements and feed-forward of the measurement results~\cite{Walgate_2000}.

Specifically, Walgate \emph{et~al.}~\cite{Walgate_2000} show that for any two orthogonal states $\ket{+}_{AB}$, $\ket{-}_{AB}$ of a bi-partite system $AB$, where for our purposes we can take $A$ to be a single qubit and $B$ an arbitrary quantum system, there always exists a decomposition of the form
\begin{equation}
\begin{aligned}\label{eq:pm orthogonal}
    \ket{+} &=  \ket{0'}_{A}\ket{\eta'_0}_B + \ket{1'}_{A}\ket{\eta'_1}_B \\
    \ket{-} &=  \ket{0'}_{A}\ket{\eta'^\perp_0}_B + \ket{1'}_A\ket{\eta'^\perp_1}_B,
\end{aligned}
\end{equation}
where $\ket{0'}$, $\ket{1'}$ are orthogonal states of $A$, and $\braket{\eta'_0}{\eta'^\perp_0} = \braket{\eta'_1}{\eta'^\perp_0} = 0$. Therefore, a measurement in the $\ket{\pm}_{AB}$ basis can be decomposed into a measurement on qubit $A$ in the $\{\ket{0'},\ket{1'}\}$ basis, feed forward of the measurement result to system $B$, followed by the simple task of distinguishing between $\ket{\eta'_i}$ and $\ket{\eta'^\perp_i}$. The procedure to construct such a decomposition is described in detail in~\cite{Walgate_2000}, and is summarised in Appendix \ref{sec:appendix}.

If $B$ is itself a single qubit, then this requires a second single-qubit measurement, conditional on the result of the first. If $B$ is multi-partite, then this decomposition can be repeated on the two orthogonal states of $B$, and so on, until we are left with a single-qubit measurement. For a measurement at level $k$ on a $(k-1)$ qubit ancilla register, this can be achieved with $(k-1)$ single-qubit measurements, with feed-forward of classical information from one measurement to the next.

\subsection{The Divide-and-Conquer Method with Disentanglement} \label{sec: disentangle}
We can now include these disentangling stages in the full description of the divide-and-conquer method for the $n=3$ case.

\begin{figure}
    \centering
    \includegraphics[width=1\linewidth]{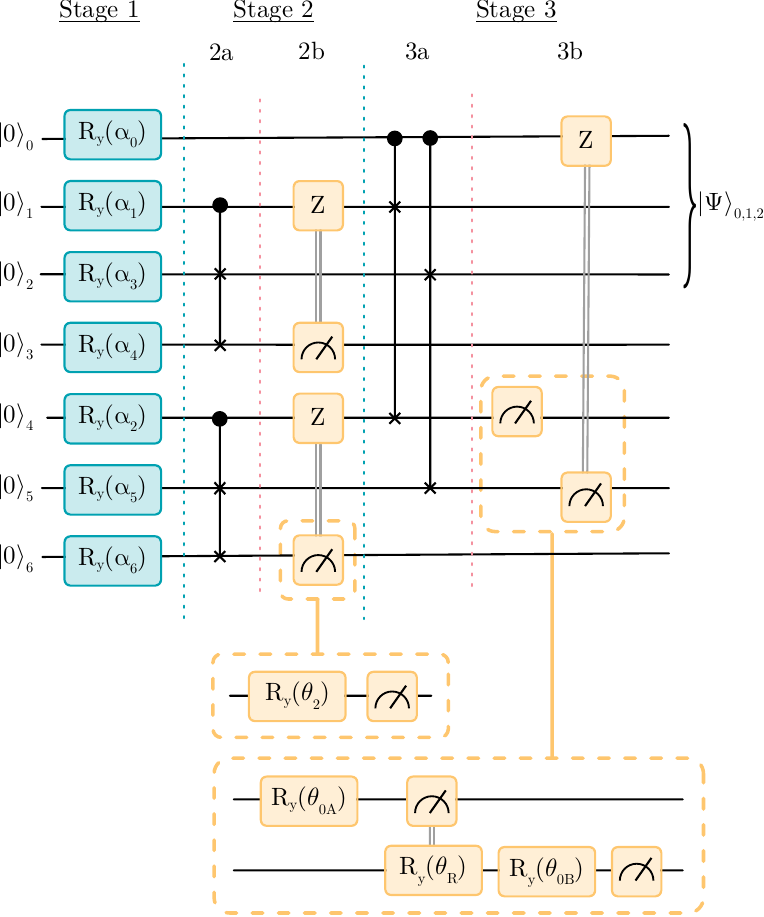}
    \caption{Quantum circuit realising the divide-and-conquer algorithm, including disentanglement stages for an N=8 input. (Double grey lines show an operation that has dependence on a classical bit.)}
    \label{fig:disentanglement}
\end{figure}
\textbf{Stage 1: 1-Qubit States}

In this paper, we use a modified order to load the rotations required in stage 1, rather than the one originally used in~\cite{Araujo_2021}. This is so that the first $n$ qubits of the circuit are always our data qubits, while the remaining are the ancilla qubits, which will be discarded.

To achieve this, we note that at each CSWAP step, the divide-and-conquer method combines two smaller binary trees, a left and a right tree. The right tree represents the qubits which are ultimately discarded, while the qubits represented by the left tree are retained. Thus, the $n$ data qubits of the final state correspond to the left-most nodes of the entire tree. By recursively implementing this at each stage of the circuit, we order the qubits in our circuit working from the root node, traversing the left subtree, then traversing the right subtree. This ordering is called the \textit{pre-order traversal} of a binary tree and will be used for the remainder of this paper. 

While this ordering is only necessary for loading the first $n$ qubits to achieve the ordered data qubit output, it also has the additional property that all simultaneous operations are drawn in parallel when generating circuit diagrams.

The first stage of the divide-and-conquer method is to load all $N-1$ single-qubit states $(\ket{\phi_0}, \ldots, \ket{\phi_6})$ into each qubit by $y$-rotations of $\alpha_0, \ldots, \alpha_6$. Using the pretraversal of the binary tree in Fig.~\ref{fig:Recursion Tree}, the order these rotations will be applied on qubits $0$ to $6$ is $(\alpha_0,\alpha_1,\alpha_3,\alpha_4,\alpha_2,\alpha_5 \text{ and }\alpha_6)$, as shown in Fig.~\ref{fig:disentanglement}.

In moving up the tree from the leaf nodes to the root, we will use the notation $\ket{\psi_{f}}$ to represent the intermediate states created at each stage. At level $l$ these denote $(n-l)$-qubit states, beginning from the first stage, $l=n-1$, where single-qubit states are represented by leaf nodes of the binary tree: $\ket{\psi_3} = \ket{\phi_3}$, $\ket{\psi_4} = \ket{\phi_4}$, $\ket{\psi_5} = \ket{\phi_5}$ and $\ket{\psi_6} = \ket{\phi_6}$.
\\
\newline
\textbf{Stage 2a: 2-Qubit States}\\
 The second stage of the divide-and-conquer method requires combining these single-qubit states pairwise to create two-qubit states. Control-SWAP operations, controlled on the states on the previous level of the binary tree, $\ket{\phi_1}$ and $\ket{\phi_2}$, create the states
\begin{equation}\label{eq:cswap1}
\begin{split}
    U_{\mathrm{CSWAP}}&(\ket{\phi_1},\ket{\phi_3},\ket{\phi_4})=\\
    & \omega_1^0\ket{0}\ket{\phi_3}\ket{\phi_4}+ \omega_1^1\ket{1}\ket{\phi_4}\ket{\phi_3},\\
    U_{\mathrm{CSWAP}}&(\ket{\phi_2},\ket{\phi_5},\ket{\phi_6})=\\
    &\omega_2^0\ket{0}\ket{\phi_5}\ket{\phi_6}+ \omega_2^1\ket{1}\ket{\phi_6}\ket{\phi_5}. 
\end{split}
\end{equation}
The desired states we wish to encode at this level are:
\begin{equation}\label{eq:psi1psi2}
\begin{split}
    \ket{\psi_1}_{1,2} = \omega_1^0\ket{0}\ket{\phi_3}+ \omega_1^1\ket{1}\ket{\phi_4},\\
    \ket{\psi_2}_{4,5} = \omega_2^0\ket{0}\ket{\phi_5}+ \omega_2^1\ket{1}\ket{\phi_6}.
\end{split}
\end{equation}
Comparing Eq.~\eqref{eq:cswap1} and Eq.~\eqref{eq:psi1psi2}, we see that there is a single-qubit ancilla remaining after the CSWAP.
\\
\newline
\textbf{Stage 2b: Disentangling Single-Qubit Registers}\\
As described in the previous subsection, disentangling the ancilla register is achieved by a measurement on the ancilla qubit in the $\ket{\pm_1}$ and $\ket{\pm_2}$ bases, respectively, where: 
\begin{align}\label{eq:pm12}
    \ket{\pm_1} = \frac{\ket{\phi_{3}} \pm \ket{\phi_{4}}}{\sqrt{2(1 \pm \braket{\phi_{3}}{\phi_{4}})}}, \hspace{0.2cm} \ket{\pm_2} = \frac{\ket{\phi_{5}} \pm \ket{\phi_{6}}}{\sqrt{2(1 \pm \braket{\phi_{5}}{\phi_{6}})}}.
\end{align}  
As is usual, we assume that all measurements are performed in the computational basis, and define the angle of rotation from the desired basis to the computational basis as $\theta_f$. Thus we perform the rotation $\theta_f$ in each case, then measure in the computational basis, achieving the desired measurements. If outcome ``1(-)" is found for either measurement, we perform a phase correctional Z-gate on the leading qubit of the corresponding state.

For this stage, the disentangling process is therefore a rotation of $\theta_1$ on qubit 3, followed by measurement, and the phase correction gate on qubit 1, to create the state $\ket{\psi_1}$ on qubits 1 and 2. Similarly, a rotation of $\theta_2$ on qubit 6, followed by measurement, and the phase correction gate on qubit 4, creates the state $\ket{\psi_2}$ in qubits 4 and 5. All steps in the preparation of $\ket{\psi_1}$, $\ket{\psi_2}$ are performed in parallel, as there are no qubits in common. This is shown in the circuit diagram in Fig.~\ref{fig:disentanglement}.\\
\newline
\textbf{Stage 3a: 3-Qubit State}\\
In the final stage, we combine the two 2-qubit states made in the previous stage to create the final 3-qubit state $\ket{\psi_0}=\ket{\Psi}$. The CSWAP is controlled on the root node of the tree $\ket{\phi_0}$. This swaps two 2-qubit registers, and is implemented using two consecutive CSWAPS, each with the same control qubit, giving:
\begin{equation}\label{eq:cswap0}
\begin{split}
    U_{CSWAP}&(\ket{\phi_0},\ket{\psi_1},\ket{\psi_2})=\\
    & \omega_0^0\ket{0}\ket{\psi_1}\ket{\psi_2}+ \omega_0^1\ket{1}\ket{\psi_2}\ket{\psi_1}.
\end{split}
\end{equation}
 The desired state at this level is:
\begin{align}
    \ket{\Psi}_{0,1,2}= \omega_0^0\ket{0}\ket{\psi_1}+ \omega_0^1\ket{1}\ket{\psi_2}.
\end{align}
In this stage, the ancilla register is a 2-qubit state, and in general, the size of the ancilla registers for each stage will be one less than the stage. \\
\newline
\textbf{Stage 3b: Disentangling 2-Qubit Registers}\\ 
The final step is to perform the disentangling measurement on the 2-qubit ancilla register. The measurement is a projector onto the two orthogonal states:
\begin{equation}\label{eq:pm0}
    \ket{\pm_0} = \frac{\ket{\psi_1} \pm \ket{\psi_2}}{\sqrt{2(1 \pm \braket{\psi_1}{\psi_2})}} \, .
\end{equation}
As described above, this can be decomposed into two quantum systems: a single-qubit $A$ (qubit 4), and the remaining qubits, $B$ (qubit 5),
\begin{equation} \label{eq:plusminusexpand}
\begin{split}
    \ket{\pm_0} = c_\pm \cdot \bigg[\ket{0}_4\big[({\omega_1^0\omega_{3}^0 \pm \omega_2^0\omega_{5}^0})\ket{0}_5 + \\(\omega_1^0\omega_{3}^1 \pm \omega_2^0\omega_{5}^1)\ket{1}_5\big]\\ +\ket{1}_4 \big[(\omega_1^1\omega_{4}^0\pm \omega_2^1\omega_{6}^0)\ket{0}_5+\\(\omega_1^1\omega_{4}^1 \pm \omega_2^1\omega_{6}^1)\ket{1}_5\big]\bigg],
\end{split}
\end{equation}
where $c_\pm$ is one over the denominator of $\ket{\pm_0}$ in Eq. (\ref{eq:pm0}).
There is a unitary matrix that transforms  Eq.~\eqref{eq:plusminusexpand} to one in the form of Eq.~\eqref{eq:pm orthogonal}.
Thus the measurement in the $\{\ket{\pm_0}\}$ basis is as follows:
\begin{enumerate}
    \item Measure qubit 4 in the $\{\ket{0'},\ket{1'}\}$ basis.\\
    \textit{Perform rotation by $\theta_{0A}$ then measure in computational basis.}
\end{enumerate}

Both $\{\ket{\eta_0'},\ket{\eta'^\perp_0}\}$ and $\{\ket{\eta_1'},\ket{\eta'^\perp_1}\}$ are orthogonal bases, so can be related by a rotation by some angle $\theta_R$, which rotates the $\{\ket{\eta_1'},\ket{\eta'^\perp_1}\}$ basis to the $\{\ket{\eta_0'},\ket{\eta'^\perp_0}\}$ basis. Thus, the measurement proceeds as follows:
\begin{enumerate}[resume]
    \item If outcome ``$0'$" is found on qubit 4, measure register 5 in the $\{\ket{\eta_0'},\ket{\eta'^\perp_0}\}$ basis.\\
    \textit{Perform rotation by $\theta_{0B}$ then measure in computational basis.}  
    \item If outcome ``$1'$" is found, measure register 5 in the $\{\ket{\eta_1'},\ket{\eta'^\perp_1}\}$ basis.\\
    \textit{Perform rotation by $\theta_{R}$ and $\theta_{0B}$ then measure in computational basis.}
\end{enumerate}
Finally, if the ``$-$" outcome is found, a correction is performed with a $Z$-gate on the leading qubit. The corresponding gates are shown in the circuit diagram in Fig.~\ref{fig:disentanglement}. In case of any combination of outcomes of measurement, the final state is given by $\ket{\Psi}_{0,1,2}$ as required.

\section{Worked Example States}
To illustrate the full divide-and-conquer algorithm, we provide examples of how this could be used in practice to prepare two 3-qubit states: a \textit{``dense"} example state and a W-state.  

\subsection{Dense State}

For a dense example, we choose a real normalised input of \\
$\vec{x}=[ \sqrt{0.04}, $$\sqrt{0.13},$$\sqrt{0.16},$$\sqrt{0.2},$$\sqrt{0.07},$$\sqrt{0.09},$$\sqrt{0.2},$$\sqrt{0.11}]$, and demonstrate how this state can be amplitude encoded using the divide-and-conquer method with the addition of a disentangling of the ancilla qubits.

Firstly, we compute the single-qubit states we wish to load in the first stage. As described previously, an input vector of size $N=8$ is represented by a binary tree with three levels, in which each node in the tree represents a single-qubit state. At the leaf nodes, using Eq.~\ref{eq:nodestate}, with $\omega_f^0, \omega_f^1$ given by Eq.~\ref{eq:omega}, these states are:

\begin{equation}
\begin{split}
   \ket{\phi_{3}}=\frac{\sqrt{0.04}\ket{0} + \sqrt{0.13} \ket{1}}{\sqrt{0.17}},
    \ket{\phi_{4}}=\frac{\sqrt{0.16}\ket{0} + \sqrt{0.2}\ket{1}}{\sqrt{0.36}}\\
   \ket{\phi_{5}}=\frac{\sqrt{0.07}\ket{0} + \sqrt{0.09} \ket{1}}{\sqrt{0.16}},
    \ket{\phi_{6}}=\frac{\sqrt{0.2}\ket{0} + \sqrt{0.11}\ket{1}}{\sqrt{0.31}}.
\end{split}
\end{equation}
Recursively applying Eq.~\ref{eq: amplitude iteration} to generate $\ket{\phi_1}$ and $\ket{\phi_2}$ produces:
\begin{equation}
   \ket{\phi_{1}}=\frac{\sqrt{0.17}\ket{0} + \sqrt{0.36} \ket{1}}{\sqrt{0.53}},
    \ket{\phi_{2}}=\frac{\sqrt{0.16}\ket{0} + \sqrt{0.31}\ket{1}}{\sqrt{0.47}}.
\end{equation}
The next and final iteration produces the state $\ket{\phi_0}$
\begin{equation}
    \ket{\phi_0}=\sqrt{0.53}\ket{0} + \sqrt{0.47}\ket{1}.
\end{equation}

Given the states $\ket{\phi_i}$ we can now define $\alpha_{f}$ from Eq.~\eqref{eq:alphaangles}. In the order that they will be loaded into the circuit, the angles used for the $y$-rotations to 2 d.p. are: $\alpha_{f}= [1.51,1.94,2.13,1.68,1.90,1.70,1.28]$.\\\newline
\textbf{Stage 1}\\
For an input of $N=8$, this will be a 7-qubit circuit. On each qubit, we perform a $y$-rotation by angle $\alpha_f$.\\
\newline
\textbf{Stage 2}
\newline
In the second stage, we create the states $\ket{\psi_1}$ and $\ket{\psi_2}$ in parallel:
\begin{equation}\label{eq:examplestate1}
\begin{aligned}
    \ket{\psi_1} &= \frac{\sqrt{0.17}}{\sqrt{0.53}}\ket{0}\ket{\phi_{3}}+ \frac{\sqrt{0.36}}{\sqrt{0.53}}\ket{1}\ket{\phi_{4}},\\
    \ket{\psi_2} &= \frac{\sqrt{0.16}}{\sqrt{0.47}}\ket{0}\ket{\phi_{5}} + \frac{\sqrt{0.31}}{\sqrt{0.47}}\ket{1}\ket{\phi_{6}}.
\end{aligned}
\end{equation}
These each require a CSWAP operation: controlled on qubit 1, with qubits 2 and 3 as the targets for $\ket{\psi_1}$; and controlled on qubit 4, with qubits 5 and 6 as the targets for $\ket{\psi_2}$. In each case, the CSWAP leaves the data qubits entangled with an ancilla qubit, and the last step in this stage is a measurement in the $\ket{\pm_1}$ basis on qubit 3 and in the $\ket{\pm_2}$ basis on qubit 6, respectively, where 
\begin{equation}
  \begin{aligned}
\ket{+_1} = 0.579 \ket{0} + 0.815\ket{1}, \\
\ket{-_1} = -0.815 \ket{0} + 0.591\ket{1}
\end{aligned}  
\end{equation}
\begin{equation}
\begin{aligned}
\ket{+_2} = 0.736 \ket{0} + 0.677\ket{1}, \\
\ket{-_2} = -0.677 \ket{0} + 0.736\ket{1}
\end{aligned}  
\end{equation}

To measure in these bases, we perform the $y$-rotation of $\theta_{1}$ then a measurement on qubit 3, and $\theta_{2}$ then a measurement on qubit 6. If the ``-/1" outcome is found, we perform a correction gate on the leading qubit of the respective state.
\\
\textbf{Stage 3}\\
Finally, in the third stage, we create state $\ket{\Psi}$:
\begin{equation} \label{eq:examplestate}
    \ket{\Psi} =\sqrt{0.53}\ket{0}\ket{\psi_1} + \sqrt{0.47}\ket{1}\ket{\psi_2}.
\end{equation}
This is achieved through CSWAPs controlled on the leading qubit, qubit 0, with target registers $\ket{\psi_1}_{1,2}$ and $\ket{\psi_2}_{4,5}$. Finally, to disentangle the ancilla register, qubits 4 and 5, we wish to measure in the basis $\ket{\pm_0}$, given in Eq.~\eqref{eq:pm0}. For this example, this becomes:

\begin{equation}\label{eq: fullpm}
\begin{split}
    \ket{+_0} = \ket{0}_4[0.332\ket{0} + 0.469\ket{1}]_5 +\\
    \ket{1}_A[0.604\ket{0} + 0.552\ket{1}]_5\\
    \\
    \ket{-_0} = \ket{0}_4[-0.534\ket{0} + 0.277\ket{1}]_5+\\
    \ket{1}_A[-0.494\ket{0} + 0.627\ket{1}]_5,
\end{split}
\end{equation}
where we have decomposed $\ket{\pm_0}$ into two subspaces, $A$ (qubit 4) and $B$ (qubit 5). Using the method of~\cite{Walgate_2000}, we can re-write these in the following form:

\begin{equation}
\begin{split}
    \ket{+_0}&=  \ket{0'}_{A}\ket{\eta'_0}_B + \ket{1'}_{A}\ket{\eta'_1}_B \\
    &=[0.833\ket{0}+0.553\ket{1}]_A[0.611\ket{0}+0.696\ket{1}]_B \\ &+[0.553\ket{0}-0.833\ket{1}]_A[-0.319\ket{0}-0.200\ket{1}]_B \\
    \\
    \ket{-_0}&=  \ket{0'}_{A}\ket{\eta_0'^\perp}_B + \ket{1'}_A\ket{\eta_1'^\perp}_B\\
    &=[0.833\ket{0}+0.553\ket{1}]_A[-0.719\ket{0}+0.578\ket{1}]_B \\ &+[0.553\ket{0}-0.833\ket{1}]_A[ 0.116\ket{0}-0.369\ket{1}]_B \\
\end{split}
\end{equation}
We then proceed as follows:
\begin{enumerate}
    \item Measure qubit 4 in the $\{\ket{0'},\ket{1'}\}$ basis.\\
    \textit{Perform rotation by $\theta_{0A}$ then measure in computational basis.}

    \item If outcome ``$0'$" is found on qubit 4, measure qubit 5 in the $\{\ket{\eta'_0},\ket{\eta'^\perp_0}\}$ basis.\\
    \textit{Perform rotation by $\theta_{0B}$ then measure in computational basis.}
    
    \item If outcome ``$1'$" is found on qubit 4, measure qubit 5 in the $\{\ket{\eta'_1},\ket{\eta'^\perp_1}\}$ basis.\\
    \textit{Perform rotation by $\theta_{R}$ and $ \theta_{0B} $ then measure in computational basis.}
\end{enumerate}

If outcome ``$-_0$" is found, apply a Z-gate on qubit 0. The final state is as given in Eq.~\eqref{eq:examplestate}. Expanding this by substituting in values for $\ket{\psi_{f}}$ at each stage, we can see that this is indeed the input state encoded in the amplitude, a 3-qubit quantum state:
\begin{equation} \label{eq:full state}
\begin{split}
    \ket{\Psi}= \sqrt{0.04}\ket{000}+\sqrt{0.13}\ket{001}+\\\sqrt{0.16} \ket{010}+\sqrt{0.2}\ket{011}+\sqrt{0.07} \ket{100}+\\\sqrt{0.09}\ket{101}+\sqrt{0.2}\ket{110}+\sqrt{0.11}\ket{111},
\end{split}
\end{equation}
as required.

\subsection{W State}
\begin{figure*}
    \centering
    \includegraphics[width=1\linewidth]{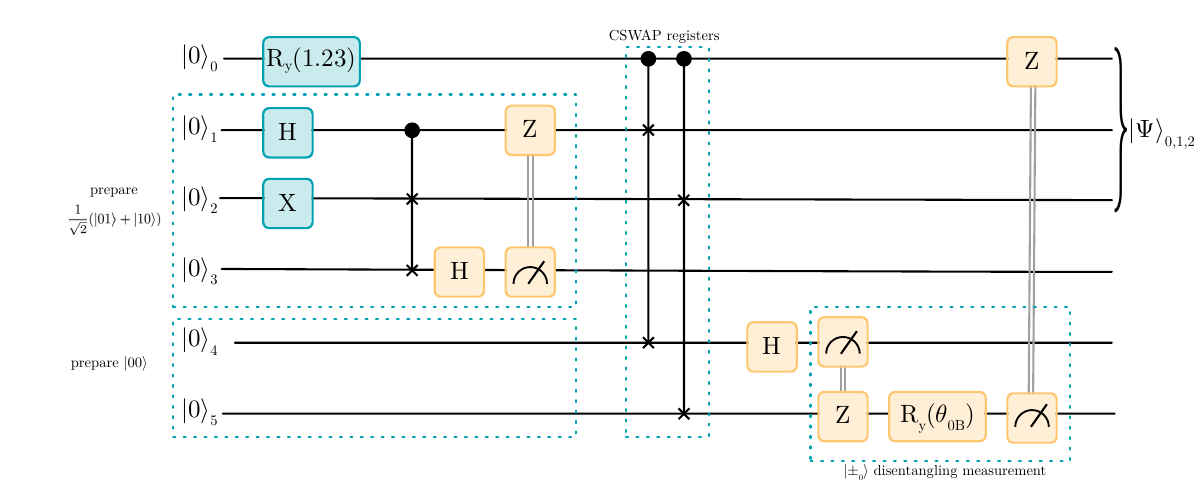}
    \caption{Quantum circuit realising the encoding of the 3-qubit W state using the divide-and-conquer method.}
    \label{fig:WState}
\end{figure*}
A commonly used class of states are the so-called W-States~\cite{D_r_2000}. 
We consider the example of a three-qubit W-State:
\begin{equation}
    \ket{W} = \frac{1}{\sqrt{3}} \left( \ket{001} + \ket{010} + \ket{100} \right).
\end{equation}
Rewriting this as:
\begin{equation}
    \ket{W}= \sqrt{\frac{2}{3}}\ket{0}_0 \frac{1}{\sqrt{2}}(\ket{0}\ket{1}+\ket{1}\ket{0} )_{1,2} + \frac{1}{\sqrt{3}}\ket{1}_0\ket{0}_1\ket{0}_2,
\end{equation}
it is clear that the two 2-qubit states we need to prepare are 

\begin{align}
\ket{\psi_1}&=\frac{1}{\sqrt{2}}(\ket{0}\ket{1}+\ket{1}\ket{0})_{1,2} \\
\ket{\psi_2}&= \ket{0}_4 \ket{0}_5.
\end{align}
In this case, no gates are needed to prepare $\ket{\psi_2}$, as all registers are initialised in $\ket{0}$. In particular, no controlled SWAP operation is needed, and therefore, no control qubit, and we can remove a qubit from the general divide-and-conquer circuit.

To prepare $\ket{\psi_1}$ we require the control qubit to be in an equal superposition of $\ket{0}$ and $\ket{1}$, and the targets in $\ket{1}$ and $\ket{0}$, achieved using a $H$, $X$, and no gate respectively.

The control-SWAP then gives:
\begin{equation}
\begin{split}
    U_{CSWAP}  (\frac{1}{\sqrt{2}}(\ket{0}+\ket{1}),\ket{1},\ket{0})=\\
    \frac{1}{\sqrt{2}}(\ket{0}\ket{1}\ket{0}+\ket{1}\ket{0}\ket{1}).
\end{split}
\end{equation}

Next, we need to disentangle the trailing qubit. It is clear that the $\ket{\pm}$ basis has the desired properties:
\begin{equation}
\ket{\pm}_1 = \ket{\pm}= \frac{1}{\sqrt{2}}(\ket{0} \pm \ket{1}).
\end{equation}
This basis achieves each outcome $\pm$ with equal probability $\frac{1}{2}$, and the residual state in each case is:
\begin{equation}
\begin{split}
    \frac{1}{\sqrt{2}}(\ket{01}_{1,2}\braket{\pm}{0}_3 \pm \ket{10}_{1,2}\braket{\pm}{1}_{3})\\
    =\frac{1}{2}(\ket{01} \pm \ket{10})_{1,2}.
\end{split}
\end{equation}

In the case that outcome ``-" is found, then a single-qubit Z operation is performed on the control qubit, qubit 1. 

The final stage is to combine the two-qubit states $(\ket{0}\ket{1}+\ket{1}\ket{0})$ and $\ket{00}$, to produce the desired three-qubit state.
To achieve the correct amplitudes in the final state, the final control qubit must be prepared in the state $\sqrt{\frac{2}{3}}\ket{0} + \sqrt{\frac{1}{3}}\ket{1}$, which is achieved through a y-rotation of $\theta_0 = 1.23$. 
Applying a controlled-SWAP operation to each pair of qubits in each register gives:
\begin{equation}
\begin{split}
    \sqrt{\frac{2}{3}}\ket{0}_0\bigg(\frac{1}{\sqrt{2}}\big(\ket{01}+\ket{10}\big)_{1,2}\bigg)\ket{00}_{4,5} \\+ 
    \sqrt{\frac{1}{3}}\ket{1}_0\big(\ket{00}\big)_{1,2}\bigg(\frac{1}{\sqrt{2}}\big(\ket{01}+\ket{10}\big)_{4,5}\bigg).
\end{split}
\end{equation}
Lastly, we need to disentangle the trailing two-qubit states from the first three data qubits. At this step, the desired measurement basis is:
\begin{equation}
    \ket{\pm_0} = \frac{1}{2}(\ket{01} + \ket{10})_{4,5} \pm \frac{1}{\sqrt{2}}\ket{00}_{4,5}.
\end{equation}

Applying the procedure in~\cite{Walgate_2000} allows us to rewrite these as:
\begin{equation}
\begin{split}
    \ket{\pm_0} = \frac{1}{\sqrt{2}}\ket{+}_4\bigg[(\frac{1}{2} \pm \frac{1}{\sqrt{2}})\ket{0}_5 + \frac{1}{2}\ket{1}_5\bigg] \\
    - \frac{1}{\sqrt{2}}\ket{-}_4\bigg[(\frac{1}{2} \mp \frac{1}{\sqrt{2}})\ket{0}_5 - \frac{1}{2}\ket{1}_5\bigg],
\end{split}
\end{equation}
where $\ket{\pm}_4 = \frac{1}{\sqrt{2}} \left( \ket{0} \pm \ket{1} \right)_4.$

Thus, the strategy is:
\begin{enumerate}
    \item Measure qubit 4 in the $\{\ket{\pm}\}$ basis.
    \item If outcome ``$+$" is obtained, measure qubit 5 in the basis $N_{\pm}[(\frac{1}{2} \pm \frac{1}{\sqrt{2}})\ket{0}_5 + \frac{1}{2}\ket{1}_5]$, where $N_{\pm}$ are constants to ensure the states are normalised.
    \item Else if outcome ``$-$" is obtained, measure qubit 5 in the basis $N_{\pm}'[(\frac{1}{2} \mp \frac{1}{\sqrt{2}})\ket{0}_5 - \frac{1}{2}\ket{1}_5]$, where again $N_{\pm}'$ are normalisation constants.
\end{enumerate}

It is useful to demonstrate this explicitly for at least one case: suppose we measure qubit 4 in  the $\{\ket{\pm}\}$ basis and obtain outcome ``+", then the residual state becomes:
\begin{equation}
\begin{split}
    \sqrt{\frac{2}{3}}\ket{0}_0\frac{1}{\sqrt{2}}(\ket{01} +\ket{10})_{1,2}\big[\frac{1}{\sqrt{2}}\ket{0}_5\big]+ \\
    \frac{1}{\sqrt{3}}\ket{100}_{0,1,2}\big[\frac{1}{2}(\ket{1} + \ket{0})_5\big].
\end{split}
\end{equation}
Finally, measuring qubit 5 in basis $\{N_\pm[(\frac{1}{2}\pm\frac{1}{\sqrt{2}})\ket{0}+ \frac{1}{2}\ket{1}]\}$ gives
\begin{equation}
\begin{split}
    N_{\pm} \sqrt{\frac{2}{3}}\ket{0}_0 \frac{1}{2}(\ket{01} + \ket{10})_{2,3}\big[\frac{1}{2} \pm \frac{1}{\sqrt{2}}\big] + \\ \frac{1}{\sqrt{3}}\ket{100}_{0,1,2} \frac{1}{2} \big[\frac{1}{2} + \frac{1}{2} \pm \frac{1}{\sqrt{2}}\big] \\
    = \frac{1}{\sqrt{2}}N_{\pm}\big[\frac{1}{2} \pm \frac{1}{\sqrt{2}}\big] \bigg(\frac{1}{\sqrt{3}}(\ket{001}+\ket{010}\pm\ket{100})_{0,1,2}\bigg)
\end{split}
\end{equation}
In the case of the ``$+$" outcome, no correction operation is needed; in the case of the ``$-$" outcome, a Z operation is applied to qubit 0. In either case, the final state upon renormalising becomes
\begin{equation}
    \frac{1}{\sqrt{3}}(\ket{001} + \ket{010} + \ket{100})\, ,
\end{equation}
as required. It is readily verified that in the case of the ``$-$" outcome on qubit 4, a similar result holds. We note that the two possible bases in qubit 5 are related by a Z operation. The full circuit is given in Fig.~\ref{fig:WState}.

Of course, there are simpler ways to prepare the W-state; however, it is instructive to service the full divide-and-conquer with a disentangling circuit for this relatively simple case as an illustration of the method.

\section{Resource requirements}
A key consideration for any state preparation procedure is to quantify the resources needed to implement it. The divide-and-conquer algorithm, in common with other methods, is designed to reduce the circuit depth, at the expense of adding ancilla qubits. Here we evaluate the circuit depth and number of ancilla needed to implement the full state preparation circuit including the disentangling steps. We also consider the classical computational cost of constructing the circuit: our method requires the calculation of the angles $ \{ \alpha_f \}$, and also the disentangling measurements $\ket{\pm}_f$. We discuss the classical computational complexity (a one-time cost) needed to construct the circuit. Finally, we consider the spacetime allocation, a measure introduced by Gui \emph{et al.}~\cite{Gui_2024}, as an analogue to the classical computing concept of ``core hours". 

For simplicity in the following analysis we assume that the time needed for single-qubit gates, two-qubit gates, and measurements is the same, and allocate one time step for each. This is not the case for most hardware platforms, and we comment on this later.

\subsection{Circuit Depth}
The initial loading stage consists of single-qubit operations, all of which may be performed in parallel, thus requiring a single time step. After this, at each stage $k$ of the algorithm, there are $(k-1)$ controlled SWAP operations. These share the same control qubit, so must be performed in sequence. The disentangling step requires $(k-1)$ single-qubit measurements. This entails feed-forward of the measurement results between qubits, and thus these must also be performed in sequence rather than in parallel. However, the CSWAP and measurement stages can overlap: for those qubits which are measured during any given stage, the measurement can be performed immediately after the CSWAP, and in parallel with other CSWAPs in the stage. There is therefore a lag of one time step between CSWAPs and measurements. Finally, there is a single-qubit correction operation, conditional on the final measurement outcome. Thus at stage $k$ there are $k+1$ operations in sequence, giving a total circuit depth of
\begin{equation}\label{eq: depth}
    d(n) = 1 + \sum_{k=2}^{n} (k+1) = \frac{1}{2}(n^2 + 3n -2).
\end{equation}
The depth of the divide-and-conquer method with measurement-based disentangling therefore scales asymptotically as $\Theta(n^2)$. Note, however, that Eq.~\ref{eq: depth} gives an exact depth calculation, which quantifies the precise resource requirements in the non-asymptotic regime, relevant for near-term applications.

\subsection{Number of qubits}
In total, the number of qubits required for the divide-and-conquer method is
\begin{equation}\label{eq: numberofqubits}
    q = \sum_{l=0}^{n-1} 2^{l} = 2^n-1,
\end{equation}
corresponding to the number of nodes in the binary tree in Figs.~\ref{fig:weighted binary tree} and~\ref{fig:Recursion Tree}, or equivalently, the number of angles $\alpha_f$. This may be further reduced if qubits can be reset and reused within the circuit. There is some discussion in the literature on mid-circuit measurements and resetting of qubits; the benefits are not just the ability to reuse a qubit, but an increase in circuit fidelity and duration~\cite{PhysRevX.13.041057,hua2023exploiting}. In the circuit described, where we disentangle the ancilla qubits from the final stage in the divide-and-conquer method, these qubits can be reused, as they are not used further on in the circuit. In particular, the first stage of our circuit requires $2^{n-1}$ leaf nodes, and $2^{n-2}$ control qubits. After this initial stage, $2^{n-2}$ qubits are measured and disentangled from the circuit. This is enough to provide those needed as controls in the next stage ($2^{n-3}$ in total). In this case the single-qubit rotation by $\alpha_f$ on these qubits is moved to just before they are first used as controls in the circuit. At all subsequent stages, there will have been enough qubits disentangled in previous stages to form control qubits without the need for any fresh qubits, thus reducing the total number of qubits needed to:
\begin{equation}
    q = 3 \times 2^{n-2}.
\end{equation}
Again this is an exact qubit count, rather than an asymptotic result.

In general, the full circuit structure outlined up to now may not be needed. A completely general 3-qubit state requires $2^3 -1 = 7$ qubits in total. However, in the W-state example discussed above, as there is only one term in the superposition in which the first qubit is in the state $\ket{1}$, we see that the resulting circuit can be implemented with six qubits. Generally, it will be possible to remove some sub-trees from the binary tree structure, depending on the particular state, leading to a corresponding circuit simplification. For sparse states where $d \ll 2^n$ non-zero components exist in the computational basis, there are $d$ non-zero amplitudes at the leaves of the tree. This implies that at the first step, we require at most $d$ single-qubit operations, and at each subsequent level, at most $d$ control qubits are needed. Overall, the encoding circuit, therefore, requires at most $nd$ qubits, which is a much more favourable scaling than the $\Theta(2^n)$ required in the completely general circuit. In general (for non-sparse states) simplifications will be possible depending on the structure of the state. We defer full discussion of this to a separate publication.

\subsection{Classical computational cost}
A general state preparation circuit is designed to encode $2^n$ classical amplitudes $\{ x_i \}$ in a quantum state, where the data is available as input to a quantum algorithm. There is always a classical computational cost associated with constructing the circuit once the data is specified. This is a one-off cost, but nevertheless must be carefully accounted for in any resource analysis. In the method presented here we need to calculate the $2^n-1$ angles $\{ \alpha_f \}$ forming the nodes of the binary tree. This is in common with other methods, and is unavoidable. Our algorithm however has an additional classical computational cost, due to the need to calculate measurement bases for the disentangling measurement, which we now address.

For a given state combining step, in which two $(k-1)$-qubit states are combined to produce one $k$-qubit state, we need to construct a $(k-1)$ qubit measurement. This is achieved using the Walgate procedure described in Appendix \ref{sec:appendix} iteratively, i.e. calculating the measurement on the first qubit, then for each possible outcome applying the same procedure to calculate the subsequent measurement on the second qubit, and so on. The Walgate procedure for two $j$-qubit states requires us to manipulate a $2 \times 2$ matrix, in which each entry is calculated as an inner product of two $2^{j-1}$ dimensional vectors, giving a cost of $\order{2^j}$.

We proceed to calculate the total classical computational cost for a $k-1$ qubit measurement: for the first qubit, we need to calculate a single measurement basis, with a cost of $\order{2^k}$. For the second qubit, we need to calculate two possible measurement bases, conditional on the result of the first measurement, each of which has a cost $\order{2^{k-1}}$. For the $j$-th qubit there are $2^{j-1}$ possible measurements, conditional on all possible previous outcomes, each with a cost of $\order{2^{k-j+1}}$. Thus for each qubit, taking into account all required calculations, the computational cost is $\order{2^k}$. There are $k-1$ qubits, giving a total cost $\order{k 2^k}$.

Finally, we need to take account of the structure of the algorithm: at stage $k$ there are $2^{n-k}$ parallel blocks, so the total number of calculations scales as $k 2^n$ at stage $k$, giving a total cost when summed over all $n-1$ levels of:
\begin{equation}
    C = \order{n^2 2^n}.
\end{equation}
Thus our method has an overhead of a factor of $n^2$ in the classical computational cost compared to the minimum.

\subsection{Spacetime allocation}
Gui \emph{et al.}~\cite{Gui_2024} introduced the metric ``spacetime allocation", as a measure of total resources needed. They define this equivalently as:
\begin{equation}
    SA := \sum_{i=0}^{q-1} d_i = \sum_{t=0}^{d-1} q_t,
\end{equation}
where $d$ is the total depth of the circuit, $q$ is the total number of qubits, $d_i$ is the total time for which the $i$-th qubit is active, and $q_t$ is the number of active qubits at layer $t$.

This is a measure of total resource cost; in particular, it is intended to quantify the cost in a situation in which multiple jobs may be scheduled on a given processor, potentially overlapping. It allows for the possibility that when a qubit is returned to the $\ket{0}$ state and not in use, it is in principle available for use in another job. 

\begin{figure}[h]
    \centering
    \includegraphics[width=0.8\linewidth]{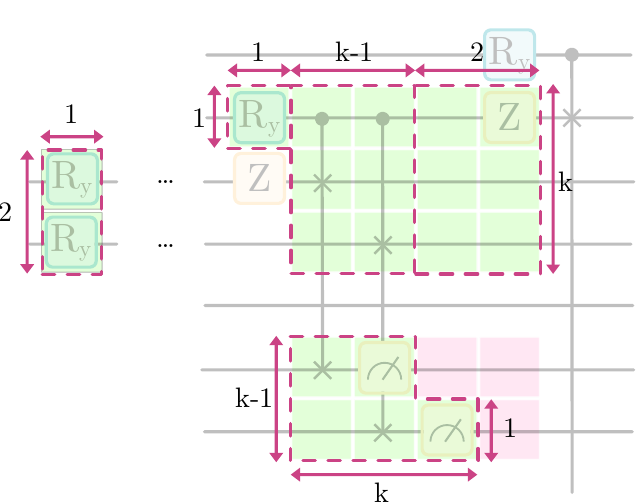}
    \caption{Example Spacetime Allocation Calculation for stage 3 of a divide-and-conquer circuit.}
    \label{fig:dac_SA_calc}
\end{figure}

In the protocol described in section \ref{sec: disentangle} we have assumed that all single rotations are performed at the start in a single time step. This means that control qubits used in later stages are left idle for multiple time steps. In the following, to reduce the idle time, we instead assume that qubits are initialised only when first used in the circuit. Thus in later stages single-qubit gates are applied directly before the first CSWAP in which a given qubit is used as a control. Similarly, we will assume that measurements are performed as soon as a qubit is no longer needed, in general in parallel with other CSWAPs at a given stage. This is shown in Fig.~\ref{fig:dac_SA_calc}.

The spacetime allocation for the total divide-and-conquer circuit can then be calculated by summing the area shown in green for each stage of the circuit. The initial loading stage of the single-qubit states representing the leaves of the tree is shown and considered separately. Subsequent stages, for $k$ from $2$ to $n$, follow the structure given, and at stage $k$ there are $2^{n-k}$ copies of each subcircuit shown.

The initial loading stage comprises $2^{n-2}$ pairs of single-qubit states, each loaded into individual qubits by rotational $y$ gates, giving a contribution of $2^{n-1}$ to the spacetime allocation:
\begin{equation}
\begin{split}
     SA_{1} &= [1\times2]\hspace{0.1cm}2^{n-2} \\
    &= 2^{n-1}\,.   
\end{split}
\end{equation}
At stage $k$, the total green area is calculated in two parts. The upper area, containing the control qubit and first target for each of $k-1$ pairs of qubits within the block, has area $k(k+1) +1$, as is readily seen from the figure.
\begin{equation}
\begin{split}
     SA_{k, upper} &= [1\times 1] + [k(k-1)] + [2k] \\
    &= k(k+1) + 1 \,.  
\end{split}
\end{equation}
For the lower area, containing the second target for each of $k-1$ pairs of qubit, we must take account of the fact that qubits are reset as soon as they are no longer needed. In the lower block, the first qubit is reset after 2 time steps, the second after 3 and so on, with the final qubit reset after $k$ time steps, as shown in Fig.~\ref{fig:dac_SA_calc}. The total area of this section is given by:
\begin{equation}
\begin{split}
     SA_{k, lower} &= 2 + 3 + \ldots k, \\
    &= \left( \sum_{j=1}^k j \right) - 1 \\
    &= \frac{1}{2} k (k+1) -1\,.
\end{split}
\end{equation}
The total spacetime allocation at stage $k$, taking into account the $2^{n-k}$ blocks is therefore:
\begin{equation}
\begin{split}
     SA_{k} &= \left( SA_{k, upper} + SA_{k, lower} \right) 2^{n-k} \\
    &= \frac{3}{2} k (k+1) 2^{n-k}.
\end{split}
\end{equation}

Summing over all stages gives the total spacetime allocation for this protocol:

\begin{equation}
    \begin{split}
        SA &= SA_1 + \sum_{k=2}^n SA_k \\
        &= 2^{n-1} + \sum_{k=1}^n \frac{3}{2} k(k+1) 2^{n-k} \\
        &= 11 \times 2^n  - \frac{3}{2} n^2- \frac{15}{2} n-12\\
        &= \order{2^{n}}.
    \end{split}
\end{equation}
This is known to be the optimal scaling for spacetime allocation~\cite{Gui_2024}.

\section{Combine and Conquer}
\begin{figure}[h!]
    \centering
    \includegraphics[width=\linewidth]{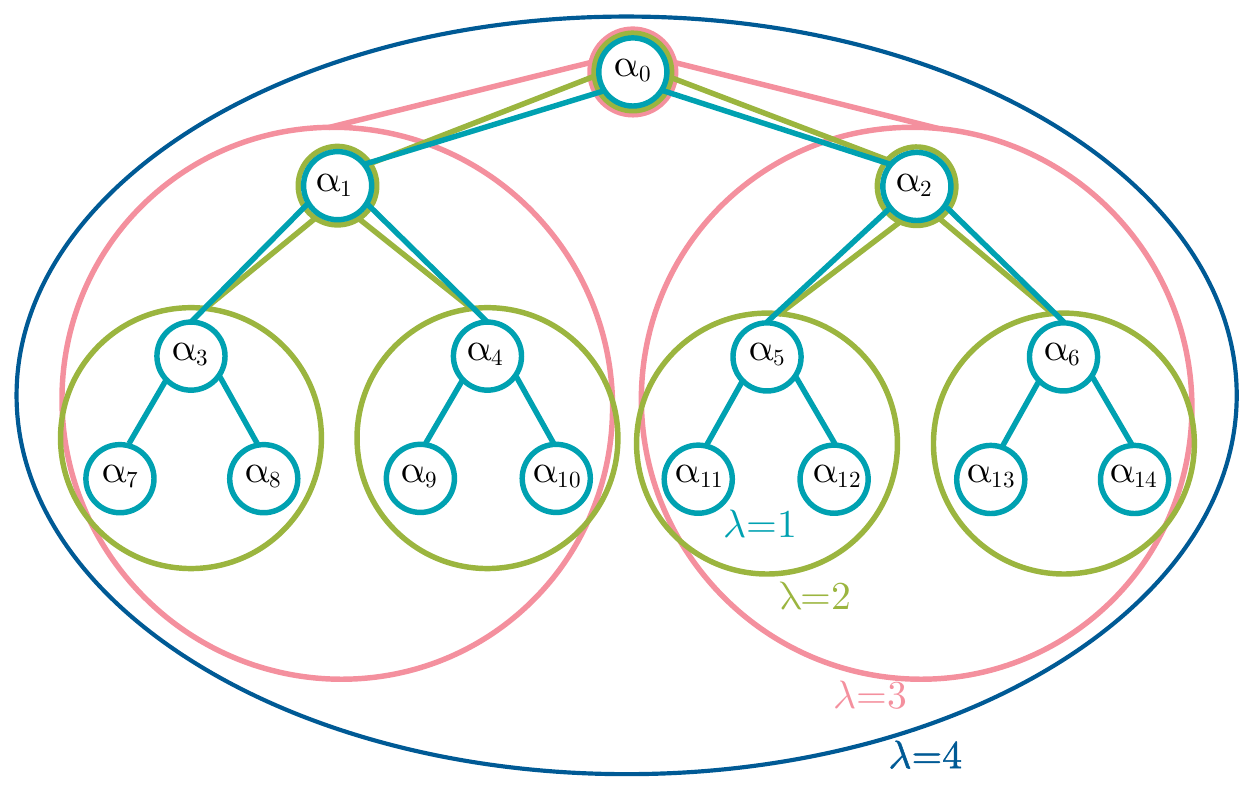}
    \caption{Grouping of sub-trees of size $\lambda$ ranging from $\lambda=1$ to $\lambda=n$ at which until level $\lambda$ an alternative encoding method has been substituted in for the divide-and-conquer method.}
    \label{fig:Combine and Conquer Tree}
\end{figure}
We note that the basic combining module, consisting of controlled-SWAPs and the disentangling measurements, is agnostic to how the states to be combined were prepared. Thus, we can readily take advantage of other methods to efficiently prepare small states corresponding to sub-trees of the overall binary tree. As an example, if an efficiently integrable function describes some part of the vector, existing methods such as the Grover-Rudolph algorithm~\cite{grover2002creating} can be used to prepare some of the smaller states efficiently. More generally, we can imagine a method which interpolates between the time encoding method and the divide-and-conquer method, using time encoding to prepare smaller $\lambda$-qubit states. This tradeoff was previously considered by~\cite{Araujo_2023}, however the final state remained entangled with an ancilla register. Our disentangling method remains applicable irrespective of how the sub-states are produced at each stage. Varying sizes of $\lambda$ are shown in Fig.~\ref{fig:Combine and Conquer Tree}.
\begin{figure}[h!]
    \centering
    \includegraphics[width=1\linewidth]{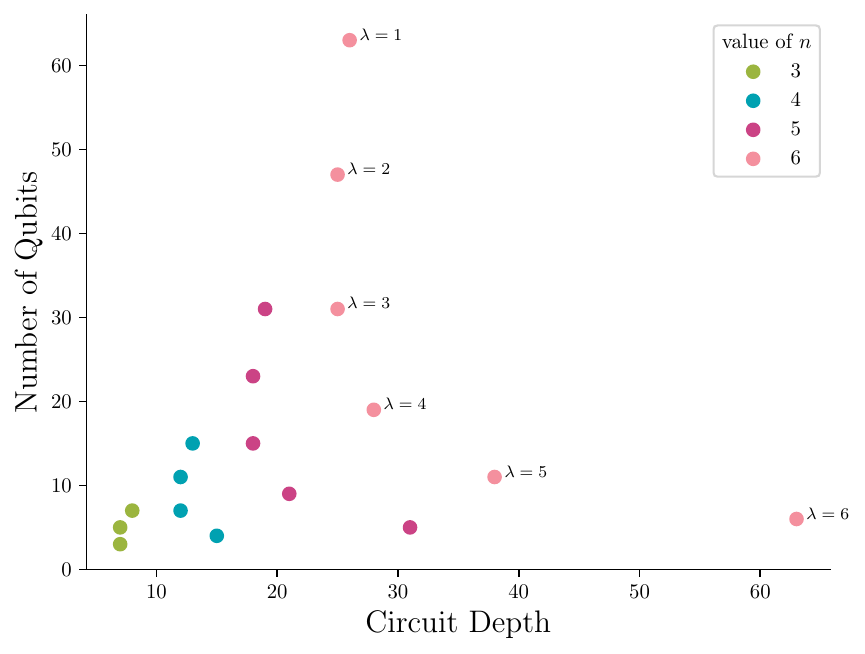}
    \caption{The relationship between the circuit depth and the number of qubits required as $\lambda$ varies from $1$ to $n$ in the combine-and-conquer circuit using time encoding.}
    \label{fig:combine and conquer plot}
\end{figure}
\newline
For the time requirements, if the time encoding method is used to create these sub-circuits, then this is an initial depth of $2^\lambda -1$, and then for each subsequent level above $\lambda$, there are $(n-l-1)$ sequences of CSWAPs; we have shown the measurement can be in conjunction with a CSWAP at each stage except the final measurement and the phase correction. Hence, the depth is
\begin{equation}
    \begin{aligned}    
    d(n,\lambda)&= 2^\lambda -1 + \sum_{l=0}^{n-\lambda-1}(n-l+1)\\
    &=  2^\lambda -1  + \frac{(n-\lambda)(n+\lambda+3)}{2} \\
    &= \order{2^\lambda + n^2}\,.
    \end{aligned}
\end{equation}
For the space requirements, the number of qubits required to create the sub-circuits is $\lambda2^{n-\lambda}$ with the remaining $2^{n-\lambda}-1$ control qubits being prepared with single-qubit rotations. Thus, the total number of qubits required is
\begin{equation}
    q(n,\lambda) = (\lambda+1)2^{n-\lambda}-1.
\end{equation}
Clearly, there is a tradeoff between circuit depth $\sim \order{2^\lambda}$ and width, or total number of qubits $\sim \order{\lambda 2^{n-\lambda}}$. For $\lambda \sim \frac{n}{2}$, these scale as $\order{2^{n/2}}$ and $\order{n2^{n/2}}$, respectively, an almost quadratic improvement over the worst scaling in the limiting cases. For $2^\lambda < n^2$ the hybrid method gives a reduction in total number of qubits needed, without increasing the scaling of the circuit depth.

 For $\lambda = 1$, the sub-circuit will prepare single-qubit rotations, reducing to the full divide-and-conquer method; for $\lambda = n$, the sub-circuit prepared is the entire circuit, so the procedure reduces to the time encoding method. When $2 \leq \lambda < n$, the circuit returned has properties from the divide-and-conquer method (acting as the base circuit) and the time encoding method (via the sub-circuits). 

 So far we have made the assumption that we can directly give up the time requirement of a circuit for additional qubits, using ancilla qubits to help prepare the state; this not only comes at the cost of using more qubits, but also the task of disentangling these qubits, adding additional depth to the circuit. We show that this added depth does not harm the overall scaling of the divide-and-conquer circuit, this is illustrated in Fig.~\ref{fig:combine and conquer plot}. We find that when combined with the combine-and-conquer circuit, we initially don't get the benefit of reduced space at the cost of increased time. We notice that when $\lambda = 1,2 \text{ or } 3$, we see the opposite effect: as larger time-encoding circuits are used, the number of qubits required decreases, but the depth also decreases. This is because, especially for small $n$, adding disentangling measurements nearly doubles the depth at that in the begining stages, as in early stages the number of measurements performed after the CSWAPs exceeds the number that can be performed in parallel. In addition, as mentioned before, the multi-control gates used in time-encoding decompose into larger gates in this approach; since we do not consider or plot further decomposition, the overhead of adding feedforward measurements to the divide-and-conquer circuits means that it initially appears to require larger depth to encode a state. This is very prominent in the $n=3$ case in Fig~\ref{fig:combine and conquer plot}, where we see the opposite behaviour than expected, this is quickly over come with larger values of $n$ as shown by  $n=5 \text{ and }6$ in the plot. The artefact of this simplistic calculation of depth is still left on the first 3 values of $\lambda$ for all $n$, but this plot should be viewed with the understanding that it is a simplified approach to quantifying the depth of a circuit and is used to get a general understanding of how the value of $\lambda$ affects the space and time requirements of a circuit.

\begin{figure}[h]
    \centering
    \includegraphics[width=1\linewidth]{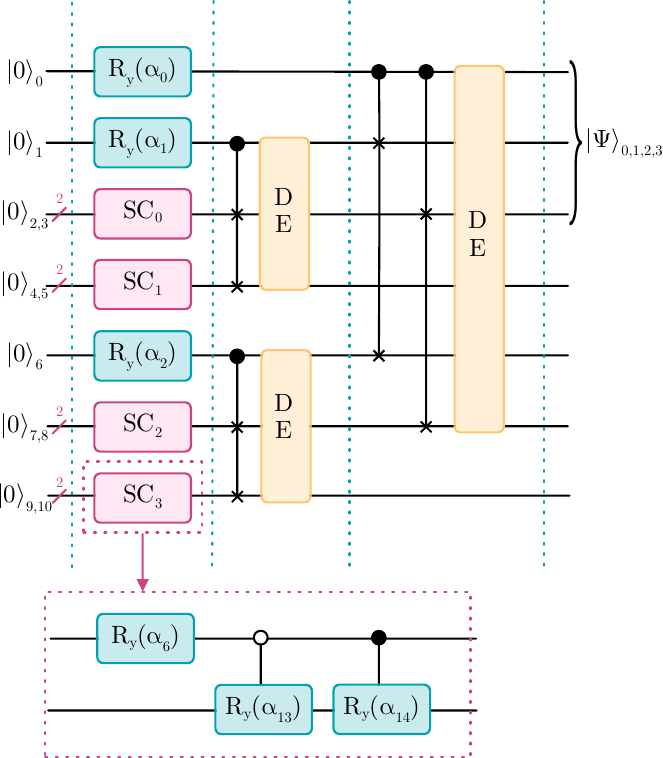}
    \caption{The general combine-and-conquer circuit for $n=4$ and $\lambda=2$}
    \label{fig: combine and conquer circuit}
\end{figure}
An example circuit is shown in Fig.~\ref{fig: combine and conquer circuit} of the binary tree from Fig.~\ref{fig:Combine and Conquer Tree}, where $\lambda = 2$. A 15-qubit circuit for the full divide-and-conquer algorithm, with 11 discarded qubits and a depth of 7 operations, becomes an 11-qubit circuit with 7 discarded qubits.

This exchange of properties between the two encoding methods can be more extensively seen in Fig.~\ref{fig:combine and conquer plot}, as a change in $\lambda$ produces a more drastic change in the circuit properties for larger values of $n$. Here, we can see explicitly the exponential nature of both methods and how the intermediary methods fit between them. We can now think of the practical implications of space and time trade-offs. For a given circuit size, we could consider distance measures on the graph to quantify the cost of each method of encoding. In Fig.~\ref{fig:combine and conquer plot}, there is a basic scaling of an additional qubit is one additional unit of space, and an additional gate is a unit of time. One could scale the axes to reflect the requirements of the hardware. For example, a circuit that prioritises low depth might be displayed on a plot with a non-linear x-axis, shifting higher depth further out, increasing the length from the origin, giving a visually intuitive way to gauge the appropriate value for $\lambda$.

\section{Conclusion}
We have introduced a measurement-based, deterministic method for state preparation of arbitrary states. This builds on the existing divide-and-conquer algorithm~\cite{Araujo_2021}, but includes disentangling of the ancilla qubits from the data qubits, resulting in a general state preparation procedure. We have precisely quantified the resources needed, in depth, number of qubits, and space-time allocation. The depth scales as $\order{n^2}$ for $\order{2^n}$ ancilla qubits, compared with an optimal depth of $\order{n}$. The spacetime allocation scales optimally as $\order{2^n}$.

The classical computational cost of this method scales as $\order{n^2 2^n}$. This is a one-off cost, which nevertheless must be accounted for. The impact of this will depend on what any subsequent processing of the data is intended to do: for classical data with $N=2^n$ data points, this means that if the subsequent data processing task can be accomplished in classical time $\order{N (\log N)^2}$ then there is no advantage to any quantum implementation using this state preparation procedure.

Our scheme is distinct from existing general state preparation protocols in that it requires mid-circuit measurements. These are less commonly used than the gate model, with the measurement at the end, but do appear elsewhere in the literature: measurement with feed-forward was previously applied to state preparation of certain classes of states in~\cite{buhrman2024statepreparation}. A potential reduction in circuit depth using dynamic circuits in other contexts has also been proposed~\cite{baumer2024quantum}. Most notably perhaps, adaptive measurements are at the heart of measurement-based quantum computing~\cite{gottesman1999demonstrating,raussendorf2001oneway}, and our suggested protocol has elements of both the gate model, and measurement-based models of quantum computing.

Dynamic circuits with this capability are increasingly available in commercial platforms, due to their central role in quantum error correction, and have been demonstrated in experiments of real-time error correction below the fault-tolerance threshold~\cite{sivak2023real,google2025quantum}. Nonetheless, the time needed for measurements is, in general, much longer than that needed for gate operations. It may be expected that these will improve in coming years, as hardware moves towards fault-tolerance. Our protocol provides a novel application of fast, high-fidelity mid-circuit measurements, and alternative motivation for their further development. In our analysis we have not distinguished between gate times for single-qubit gates, multi-qubit gates, and measurements. The detailed implementation in particular hardware, each with its own connectivity and specifications is left for future work. One could imagine a state preparation module in which CSWAP gates and adaptive measurements are optimised.

There are now a number of different general state preparation protocols with optimal or near-optimal scaling in various resource quantifiers. In the short to medium term, with limited quantum resources, it will be the precise resource requirement which is most relevant, and not just the scaling. There is a need to better understand the practical applications of each method, including a careful comparison of the performance of each method on different platforms.

Our state preparation method is versatile, and we have described how this can be combined with existing methods to generate a range of protocols with different resource tradeoffs. We note that although the circuits may have exponentially many qubits, due to the modular nature of the protocol entanglement is always confined to subsets of at most $n$ qubits. We will explore further modifications of this basic protocol, with varying resource requirements, in a separate paper. The example of the W-state shows that in some cases, not all qubits and gates are needed. We will also explore elsewhere the optimisation of the general structure of the circuit to reduce the overall number of qubits needed, tailored to the particular target state.

\section*{acknolwedgements}
This work was supported by the University of Glasgow via the James McCune Smith Scholarship.
This work was also supported by the UK Engineering and Physical Sciences Research Council (EP/Y014456/1).

\section*{Code Availablity}

Code to create divide-and-conquer/ combine-and-conquer circuits as described in the paper can be found at: \newline
\url{https://github.com/Roselyn-Nmaju/Low-depth-measurement-based-deterministic-quantum-state-preparation.git}.
\newpage
\appendix
\section{Local adaptive measurement procedure \label{sec:appendix}}
Here we follow closely the discussion of~\cite{Walgate_2000}, summarising the construction given of a decomposition of the form Eq.~\ref{eq:pm orthogonal} for two arbitrary orthogonal states $\ket{\pm}_{AB}$. We consider $AB$ to be composed of two quantum systems: a single-qubit $A$ and the remaining qubits $B$, such that
\begin{equation}
\begin{aligned}
    \ket{+} &=  \ket{0}_A\ket{\eta_0}_B + \ket{1}_A\ket{\eta_1}_B, \\
    \ket{-} &=  \ket{0}_A\ket{\nu_0}_B + \ket{1}_A\ket{\nu_1}_B,
\end{aligned}
\end{equation}
where $\ket{0}_A$, $\ket{1}_A$ are orthogonal states of $A$, while $\ket{\eta_i}$ and $\ket{\nu_i}$ are neither orthogonal nor normalised in general. The key result of~\cite{Walgate_2000} is that by way of a unitary transformation of $U^A$, we can always find a basis $\{\ket{0'},\ket{1'}\}$ of $A$ such that $\ket{\pm}_{AB}$ can be written as:
\begin{equation}
\begin{aligned}\label{eq:pm appendix}
    \ket{+}_{AB} &=  \ket{0'}_{A}\ket{\eta'_0}_B + \ket{1'}_{A}\ket{\eta'_1}_B \\
    \ket{-}_{AB} &=  \ket{0'}_{A}\ket{\eta_0'^\perp}_B + \ket{1'}_A\ket{\eta_1'^\perp}_B.
\end{aligned}
\end{equation}
Therefore, the measurement can be decomposed into two steps. In the first step a measurement is performed on qubit $A$ in the $\{\ket{0'},\ket{1'}\}$ basis. Then, the measurement result is fed forward to system $B$, where the simple task of distinguishing between $\ket{\eta'_i}$ and $\ket{\eta_i'^\perp}$ remains.

We outline the construction of $U^A$ for the case in which $B$ is a single qubit, and refer the reader to~\cite{Walgate_2000} for the more general case. We recognise that $\ket{\eta_i}$ and $\ket{\nu_i}$ can be expressed as a superposition of basis vectors, $ \{\ket{0}, \ket{1} \}$:
\begin{align}
    \ket{\eta_i}_B = \sum_j F_{ij}\ket{j}_B, \quad \ket{\nu_i}_B = \sum_j G_{ij}\ket{j}_B,
 \end{align}
where the elements of $F_{ij}$ and $G_{ij}$ preserve the information of the $\{\ket{\pm}\}$ basis, so we can construct the matrix $FG^\dag$:
\begin{equation}
    FG^\dag = \begin{bmatrix} \braket{\nu_0}{\eta_0} & \braket{\nu_0}{\eta_1} \\ \\ \braket{\nu_1}{\eta_0} & \braket{\nu_1}{\eta_1} \end{bmatrix}.
\end{equation}
By the construction of the $\{\ket{\pm}_{AB}\}$ basis it must hold that,
\begin{equation}
    \braket{+}{-} = \sum_{i=0}^1 \braket{\nu_i}{\eta_i} = \textrm{Trace}(FG^\dag) = 0.
\end{equation}
This matrix holds the key to distinguishing between our two states; via a unitary transformation on $A$ $(U^A)$, we can find a basis such that we can write the  $\ket{\pm}$ in the form of Eq.~\eqref{eq:pm orthogonal}. Under transformation $(U^A)$
\begin{equation}
    FG^\dag \rightarrow U^A(FG^\dag)U^{A\dag} =F'G'^\dag.
\end{equation}
In this new basis, $F'G'^\dag$ is also a matrix which preserves the elements of the $\{\ket{\pm}\}$ basis but with all of its diagonal entries equal to 0, so it must hold that under the basis transform of $(U^A)$, 
\begin{equation}
    \ket{\eta_i} \rightarrow\ket{\eta'_i}, \quad \ket{\nu_i} \rightarrow\ket{\nu'_i}=\ket{\eta'^\perp_i}.
\end{equation}
$U^A$ takes the general form:
\begin{equation}
    U^A = \begin{bmatrix} \cos(\theta) & \sin(\theta)e^{i\omega} \\ \\ \sin(\theta)e^{-i\omega} & -\cos(\theta) \end{bmatrix},
\end{equation}
where we use the notation of~\cite{Walgate_2000}, and parameters $\theta$, $\omega$ should not be confused with those used in the main body of our paper. Enforcing zeros in the diagonal elements of $F'G'^\dag$ gives the constraint
\begin{equation}
\begin{split}
    (\braket{\nu_0}{\eta_0}- \braket{\nu_1}{\eta_1} )\cos{(2\theta)} &+\\ (\braket{\nu_0}{\eta_1}e^{-i\omega}+\braket{\nu_1}{\eta_0}e^{i\omega})\sin{(2\theta)}&=0.
\end{split}
\end{equation}
From this, we solve for the real and imaginary parts of $\omega$ and $\theta$:

From this, we solve for the real and imaginary parts of $\omega$ and $\theta$:
\begin{equation}
\begin{aligned}
    a      = \braket{\nu_0}{\eta_0} - \braket{\nu_1}{\eta_1}, \\
    b_\pm  = \braket{\nu_1}{\eta_0} \pm \braket{\nu_0}{\eta_1}.
\end{aligned}
\end{equation}
\begin{equation}
        \tan(\omega) = \frac{\Im{a}\,\Re{b_+} - \Re{a}\,\Im{b_+}}
                         {\Re{a}\,\Re{b_-} + \Im{a}\,\Im{b_-}}.
\end{equation}
\begin{align}
    \tan(2\theta)= \frac{a}{\Re{b_+} \cos(\omega)-\Im{b_- }\sin(\omega)}.
\end{align}
The two-qubit measurement $\ket{\pm}$ becomes a measurement in the $\{\ket{0'},\ket{1'}\}$ basis on qubit $A$, whose result is fed forward to determine the distinguishing measurement performed on register $B$.
If $B$ is multi-partite, this process can be repeated on the two orthogonal $B$ states until qubit $B$ is a single-qubit state.

For a measurement of $k-1$ qubits, this can be achieved with $(k-1)$ single-qubit measurements, with $(k-2)$ of those measurement results used to condition the subsequent measurements.

\newpage
\bibliography{References}

\end{document}